\documentclass[%
 reprint,
 amsmath,amssymb,
 aps,
]{revtex4-2}

  \makeatletter
\def\@fnsymbol#1{\ensuremath{\ifcase#1\or \dagger\or \ddagger\or
   \mathsection\or \mathparagraph\or \|\or **\or \dagger\dagger
   \or \ddagger\ddagger \else\@ctrerr\fi}}
    \makeatother

\makeatletter

\def\blfootnote{\gdef\@thefnmark{}\@footnotetext}

\makeatother

\usepackage{graphicx}% Include figure files
\usepackage{dcolumn}% Align table columns on decimal point
\usepackage{bm}% bold math
\usepackage{xcolor}
\usepackage[normalem]{ulem}
\begin{document}

\preprint{APS/123-QED}

\title{Reverse Intersystem Crossing Dynamics in Vibronically Modulated Inverted Singlet-Triplet Gap System: A Wigner Phase Space Study }% Force line breaks with \\
\blfootnote{A footnote to the article title}%

\author{Pijush Karak}
\altaffiliation{Department of Chemistry, University of Calcutta, 92 A. P. C. Road,
Kolkata–700009, West Bengal, India.}
\author{Pradipta Manna}
\altaffiliation{Department of Chemistry, University of Calcutta, 92 A. P. C. Road,
Kolkata–700009, West Bengal, India.}
\author{Ambar Banerjee$^{*}$}
\altaffiliation{Research Institute for Sustainable Energy (RISE), TCG Centres for Research and Education in Science and Technology (TCG-CREST), Kolkata, 700091 India
}
\email{ambar.banerjee@tcgcrest.org.}

\author{Kenneth Ruud$^{*}$}
\altaffiliation{Hylleraas Centre for Quantum Molecular Sciences, Department of Chemistry,
University of Tromsø – The Arctic University of Norway, 9037 Tromsø, Norway.}
\altaffiliation{Norwegian Defence Research Establishment, P.O.Box 25, 2027 Kjeller, Norway}
\email{kenneth.ruud@uit.no.}

\author{Swapan Chakrabarti$^{*}$}
\altaffiliation{Department of Chemistry, University of Calcutta, 92 A. P. C. Road, Kolkata–700009, West Bengal, India}
\email{swcchem@caluniv.ac.in.}


\date{\today}% It is always \today, today,

\begin{abstract}
We inspect the origin of the inverted singlet-triplet gap (INVEST) and slow change in the reverse intersystem crossing (rISC) rate with temperature, as recently observed. A Wigner phase space study reveals, that though INVEST is found at equilibrium geometry, variation in the exchange interaction and the doubles-excitation for other geometries in the harmonic region leads to non-INVEST behavior. This highlights the importance of nuclear degrees of freedom for the INVEST phenomenon and in this case, geometric puckering of the studied molecule determines INVEST and the associated rISC dynamics.

\end{abstract}

\maketitle

%\tableofcontents

The breakdown of a rule of nature usually helps us understand the rule in a better way and often appears as a blessing in the advancement of new technologies. One such rule is the notion that the triplet ground state (T$_{1}$) is more stable than the first excited singlet state (S$_{1}$), as predicted by Hund's rule\cite{Hund_1925}. In 2019, Domcke and co-workers\cite{Ehrmaier_2019_jpca_s_t_inversion_heptazine_polymeric_carbon_nitride} showed that this rule can be broken in a class of structurally rigid heptazine molecules. It has been argued that reduced exchange interaction due to poor overlap between the relevant orbitals of the T$_{1}$ state and concomitant additional stabilization of the S$_{1}$ state caused by doubles-excitation contributions lead to the inversion of the singlet-triplet gap and thus the breakdown of Hund's rule. This class of inverted singlet-triplet gap molecules is popularly known as INVEST\cite{Aizawa_2022_nature_invest_delayed_fluorescence,Ehrmaier_2019_jpca_s_t_inversion_heptazine_polymeric_carbon_nitride,Sobolewski_2021_jpcl_heptazine_oled_tadf_inverted_s_t_gap,Omar_2023_jacs_inveretd_s_t_high_thoughtput_screening} and  are now on the centre stage of delayed fluorescence\cite{Sato_prl_2013_zero_energy_gap_efficient_tadf,LIn_2003_prl_triplet_singlet_exciton_poly_vinylene_oled,Police_2021_matter_inverted_s_t_gap_fluorescence_rate,Eungdo_2022_sci_adv_tadf,Elad_2019_sci_adv_invest_strong_light_mater_coupling_delayed_fluo,Yu_2021_nat_commun_barrier_free_risc} research because thermally promoted reverse intersystem crossing\cite{Talotta_2020_prl_spin_orbit_effect_ultrafast_molecular_process,Chen_2024_pra_tadf_oled,Hasan_2022_pra_tadf_oled,Vander_2022_pra_oled_tadf} between T$_{1}$ and S$_{1}$ has become barrier-free energy saving process. A significant amount of work has been done both theoretically\cite{Garner_2023_chem_sci_double_bond_delocalization_non_alternate_hydrocarbon_inverted_s_t_gap,Bedogni_2024_jctc_inverted_singlet_triplet_emitter,de_silva_2019_jpcl_inverted_s_t_gap_relevance_tadf} and experimentally\cite{Ehrmaier_2019_jpca_s_t_inversion_heptazine_polymeric_carbon_nitride,Aizawa_2022_nature_invest_delayed_fluorescence,Blasco_2024_pccp_expt_comput_optical_properties_heptazine_derivative_s_t_gap} over the last three to four years. It has been found at the experimental level that 2,5,8-tris(4-
methoxyphenyl)-1,3,5,6,9,9b-heptaazaphenalene 2,5,8-tris(4-fluoro-3-methylphenyl)heptazine or tri-anisole-heptazine exhibit INVEST.\cite{Rabe_jpcl_2019_heptazine_expt,Ehrmaier_2019_jpca_s_t_inversion_heptazine_polymeric_carbon_nitride}
At the same time, various wavefunction-based quantum-chemical methods\cite{Drew_2023_front_chem_inverted_singlet_triplet_gap,Loos_2023_jpcl_heptazine_cyclazine_chemically_acurate_emitter_inverted_s_t_gap,Sanz-Rodrigo_2021_jpca_s_t_gap_traingle_shaped_molecule_triplet_harvesting,Drawl_jctc_2023_spin_polarization_dynamic_corr_s_t_gap_heptazine,Sobolewski_2023_pccp_s_t_gap_hexagonal_aromatic_heteroaromatic_compounds} and doubles-corrected double hybrid functionals in the framework of time-dependent density functional theory(TDDFT)\cite{Sancho-Garcia_2022_jcp_violation_Hund_rule_excited_state_energy_inversion_tddft_double_hybrid_methods,ALipour_2022_jcp_whether_double_hybrid_render_correct_order_s_t_state_energy_gap} for the prediction of INVEST has been extensively benchmarked. It is also worth noting that INVEST can also be achieved by creating a polariton\cite{Joshua_2024_nat_commun_polariton,Fan_2023_prl_polariton,Ruggenthaler_2023_chem_rev_polariton,Bhuyan_2023_chem_rev_polariton,Elad_2019_sci_adv_invest_strong_light_mater_coupling_delayed_fluo,Flick_2018_prl_cavity_elctron_nulcear_dynamics,Memmi_2017_prl_polariton,Lopez_2015_prl_polariton} in an optical cavity\cite{Ashida_2021_prl_cavity_QED,Yu_2021_nat_commun_barrier_free_risc,Kati_2018_nat_commun_polariton,Qi_jacs_2021_polariton} setup. \\

\begin{figure*}
\includegraphics{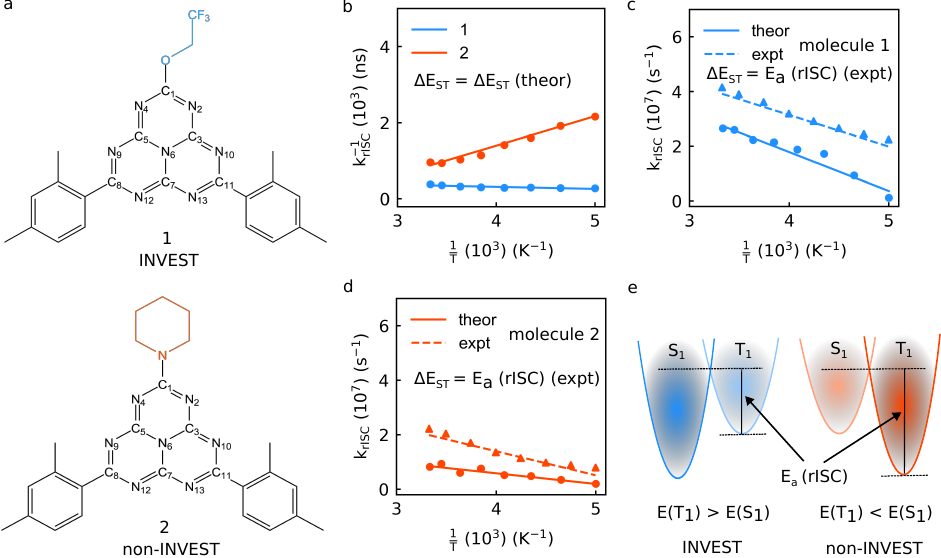}
\caption{\label{fig:molecular-structures-temp}(a) Chemical structures of INVEST (\textbf{1}) and non-INVEST (\textbf{2}) molecules. (b) Variation of $\mathrm{k_{rISC}^{-1}}$ with temperature (T) of INVEST (\textbf{1}) and non-INVEST (\textbf{2}) molecules considering the theoretically computed singlet-triplet energy gap. The change in rISC rate constant represented in (c) and (d) of molecules \textbf{1} and \textbf{2} ,respectively, are obtained by replacing the energy gap parameter ($\mathrm{\Delta E_{ST}}$) of Eq.~(\ref{eqn:kisc-gdso-part}) with the experimental activation energy ($\mathrm{E_{a}}$). The experimental value of $\mathrm{E_{a}}$ of INVEST and non-INVEST are 0.042 and 0.069 eV, respectively. The behavior of the experimental and the theoretical k$_{\mathrm{rISC}}$ with temperature are denoted by solid triangle and circle, respectively. The dashed and solid lines indicate the least square fitting of the data points for experimental and theoretical results, respectively. (e) The activation energy ($\mathrm{E_{a}}$) of the rISC mechanism is shown pictorially for both INVEST and non-INVEST molecules. Here $\mathrm{E(S_{1})}$ and $\mathrm{E(T_{1})}$ imply the energy of the lowest excited singlet (S$_{1}$) and the triplet (T$_{1}$) states, respectively.}
\end{figure*}

In the most illustrious experiment done so far, a lowering of the delayed fluorescence (DF) lifetime with a decrease in temperature was observed\cite{Aizawa_2022_nature_invest_delayed_fluorescence},  and subsequently this INVEST molecule was used in the fabrication of OLEDs\cite{Uoyama_2012_nature_organic_light_emit_diode_df,Zhang_prl_2012_oled,Giebnik_prl_2023_oled,Braun_2023_pra_upconversion_triplet_singlet_fluorescence_phosphorecence_oled,Chen_2024_pra_tadf_oled,Won_2023_chem_phys_rev_inverted_s_t_gap_oled}.  In Ref.~\cite{Aizawa_2022_nature_invest_delayed_fluorescence}, the authors initially performed a computational screening of 34,596 substituted heptazine molecules  and after step-wise elimination, they were in the end able to synthesize a few molecules for experimental analysis, of which, 2-(2,2,2-trifluoroethoxy)-
5,8-bis(2,4-xylyl)heptazine (see FIG.~\ref{fig:molecular-structures-temp}a) and two more systems exhibited INVEST. Nonetheless, the questions of how INVEST responds to vibronic interactions at finite temperatures and what drives the slow lowering of the DF lifetime of INVEST molecules with lowering of the temperature,  remain unanswered.

To address these fundamental questions, in this letter, we perform computations on the temperature-dependent rate constant of rISC for both INVEST and non-INVEST molecules using a generating function-based, in-house  code.\cite{swapanchem_risc} Moreover, to gain insight into the vibronic effects on INVEST/non-INVEST at finite temperature, we have conducted Wigner phase space sampling using the SHARC software\cite{sebastian_mai_2023_sharc_software,Mai_2018_WIRES_SHARC_nonadiabatic_dynamics} to generate 300 geometries of the studied systems and computed the gap between S$_{1}$ and T$_{1}$ along with the decomposition of the singlet-triplet gap into contributions from doubles excitations and exchange interactions. Our study reveals that normal-mode vibrations along three dihedral angles involving central and peripheral nitrogen atoms of the heptazine moiety exhibiting out-of-plane motion of the nitrogen and the carbon atoms have profound impact on the modulation of the energy gap of INVEST. Furthermore, the thermal rate constants of rISC in these selected dihedral angles show slow variation of the rate constants with temperature.

To compute the rate constant of ISC and rISC between two vibronic states, we have used a generating function-based method in the framework of Fermi's golden rule and the basic equation used in our already developed code is 
\begin{eqnarray}
\mathrm{k_{rISC}}&&=\frac{2\pi}{\hbar}\sum_{v_{\mathrm{S_{1}}},v_{\mathrm{T_{1}}}}\frac{e^{-\beta E_{v_{\mathrm{T_{1}}}}}}{z} 
  \lvert \langle {\mathrm{T_{1}},v_{\mathrm{T_{1}}} \lvert \hat{H}_{\mathrm{SO}} \rvert \mathrm{S_{1}},v_{\mathrm{S_{1}}}} \rangle \rvert ^{2} \nonumber \\
 && \times \delta (-\Delta E _{\mathrm{ST}}+E_{v_{\mathrm{S_{1}}}}-E_{v_{\mathrm{T_{1}}}})~.
 \label{eqn:krisc-dso-part}
\end{eqnarray}
In the above equation, the singlet and triplet vibronic states are denoted by $\lvert \mathrm{S_{1}},v_{\mathrm{S_{1}}}\rangle$ and $\lvert \mathrm{T_{1}},v_{\mathrm{T_{1}}}\rangle$, respectively. E$_{v_{\mathrm{S_{1}}}}$ and E$_{v_{\mathrm{T_{1}}}}$ implies the energy of the singlet and triplet vibrational states, respectively. $\Delta E _{\mathrm{ST}}$ and $z$ represents the singlet-triplet energy gap and partition function of the initial electronic state, respectively. The effect of temperature in Eq.~(\ref{eqn:krisc-dso-part}) has been included by $\sum_{v_{\mathrm{T_{1}}}}\frac{e^{-\beta E_{v_{\mathrm{T_{1}}}}}}{z}$, where $\beta=\frac{1}{kT}$, with $k$ and $T$ are denoting the Boltzmann constant and temperature, respectively.  
The final form of the rISC rate equation in terms of the time-dependent generating function is expressed as
\begin{eqnarray}
{\mathrm{k_{rISC}}}= \frac{\vert {{H}}_{\mathrm{SO}}\vert^{2}}{\hbar^{2} z}\int_{-\infty}^{\infty}G(t,t^\prime)e^{-i\Delta E_{\mathrm{ST}}\frac{t}{\hbar}} dt,
\label{eqn:kisc-gdso-part}
\end{eqnarray}
where $G(t,t^\prime)$ is the time-dependent generating function and the relation between $t$ and $t^{\prime}$ is $t^{\prime}=-\beta-\frac{it}{\hbar}$. The detailed derivation of Eq.~(\ref{eqn:kisc-gdso-part}) can be found in Refs.\cite{Kim_jctc_2020_spin_vibronic_risc_tadf_solvent,Pijush_jcp_2022_spin_vibronic_risc,KARAK2023}.
The basic input variables used to perform the rate constant computations are the energy gap between the S$_{1}$ and T$_{1}$ states, spin--orbit coupling, Duschinsky rotation matrix, displacement vectors and the normal-mode frequencies of the mentioned states. For calculating the energy gap, we have used a double hybrid functional which includes doubles corrections, namely $\omega$B2GP-PLYP, which has been extensively tested and verified to correctly predict the inverted singlet-triplet gap  for experimentally observed INVEST phenomenon in 2,5,8-tris(4-fluoro-3-methylphenyl)heptazine (HAP-3MF), tri-anisole-heptazine (TAHz) and 2-(2,2,2-trifluoroethoxy)-
5,8-bis(2,4-xylyl)heptazine (\textbf{1})\cite{ALipour_2022_jcp_whether_double_hybrid_render_correct_order_s_t_state_energy_gap}. The computational details for the other variables are provided in the supplementary file.

\begin{figure*}
\includegraphics{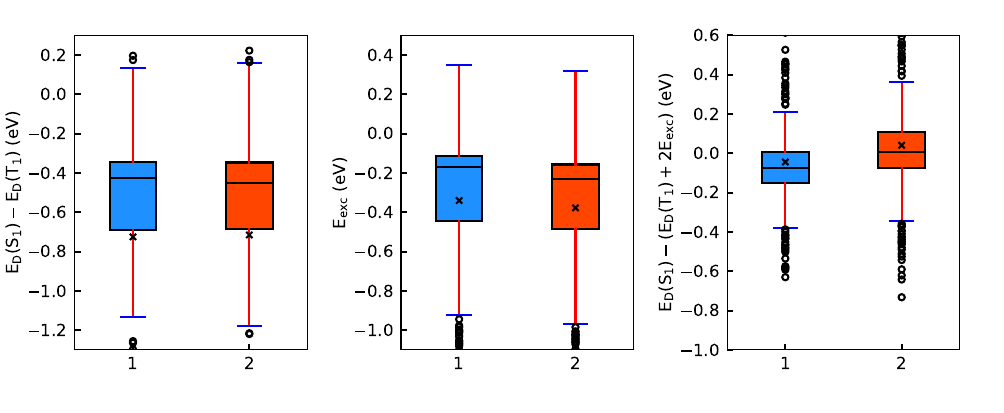}
\caption{\label{fig:box-whisker-all}Box-whisker plot of energy change of the S$_{1}$ state compared to that of the T$_{1}$ state due to doubles excitation configuration (left), exchange energy (middle) and net effect of doubles and exchange energy (right) computed on Wigner distributed 300 geometries at 300 K of molecules \textbf{1} and \textbf{2}. E$_{\mathrm{D}}(\mathrm{S_{1}})$ and E$_{\mathrm{D}}(\mathrm{T_{1}})$ are doubles corrected energies of S$_{1}$ and T$_{1}$ states, respectively. E$_{\mathrm{exc}}$ indicates the exchange energy. The median and mean value are denoted by black solid line and cross point, respectively. }
\end{figure*}

\begin{figure*}
\includegraphics{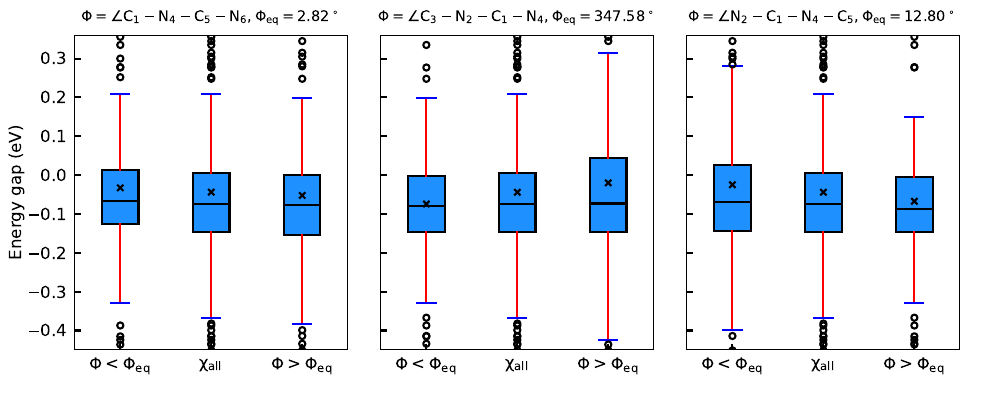}
\caption{\label{fig:box-whisker-plot-three-selected-da-300K}Distribution of the energy gap between the S$_{1}$ and the T$_{1}$ state of an ensemble containing Wigner distributed geometries at 300 K of three selected dihedral angles, namely, $\angle \mathrm{C_{1}-N_{4}-C_{5}-N_{6}}$ (left), $\angle \mathrm{C_{3}-N_{2}-C_{1}-N_{4}}$ (middle) and $\angle \mathrm{N_{2}-C_{1}-N_{4}-C_{5}}$ (right) are shown by Box-whisker plots of molecule \textbf{1}. Here, $\mathrm{\chi_{all}}$ denotes the inclusion of all 300 geometries. $\mathrm{\Phi}<\mathrm{\Phi_{eq}}$ and $\mathrm{\Phi}>\mathrm{\Phi_{eq}}$ are suggesting the geometries having dihedral angle less and greater, respectively, than their corresponding equilibrium values ($\mathrm{\Phi_{eq}}$). }
\end{figure*}

Delayed fluorescence (DF) is a complex phenomenon and the associated lifetime depends on many factors, such as the rate of ISC/rISC as well as radiative and internal conversion decay rates from the S$_{1}$ state. Our analysis of these different factors have revealed a correlation between the experimental temperature-dependent DF lifetime with the calculated k$_{\mathrm{rISC}}^{-1}$. The temperature-dependent k$_{\mathrm{rISC}}^{-1}$ both for the molecule exhibiting INVEST and that exhibiting non-INVEST are presented in FIG.~\ref{fig:molecular-structures-temp}b. It is evident from FIG.~\ref{fig:molecular-structures-temp}b that k$_{\mathrm{rISC}}^{-1}$ of molecule \textbf{1} decreases with temperature while for molecule \textbf{2},  the reverse trend is observed. The faster rate of delayed fluorescence due to the downhill rISC mechanism in molecule \textbf{1} probably leads to the shorter delayed fluorescence lifetime ($\tau_{\mathrm{DF}}$) at low temperature. 
The computed k$_{\mathrm{rISC}}$ values between the T$_{1}$ and the S$_{1}$ states of molecules \textbf{1} and \textbf{2} are 2.6$\times 10^{7}$ and 1.0$\times 10^{7}$ s$^{-1}$, respectively. These theoretical results are in good agreement with the experimentally\cite{Aizawa_2022_nature_invest_delayed_fluorescence} observed rISC rate constants of  4.2$\times 10^{7}$ s$^{-1}$ and 2.2$\times 10^{7}$ s$^{-1}$ for molecules \textbf{1} and \textbf{2}, respectively.
The experimental\cite{Aizawa_2022_nature_invest_delayed_fluorescence} $\Delta E_\mathrm{{ST}}$ values of molecule \textbf{1} and molecule \textbf{2} are -0.011 and 0.052 eV, respectively, and these values agree well with the computed gap of -0.089 and 0.01 eV, respectively. The small gap irrespective of the INVEST and non-INVEST nature of the studied molecules necessitates an analysis of the role of molecular vibrations on the electronic energy gap and their complex effect on the temperature dependence of the rISC processes. 

To this end, we have performed  Wigner phase space sampling of molecule \textbf{1} and \textbf{2} at 0 and 300 K to generate an ensemble of 300 geometries in the harmonic region where the bond lengths, bond angles and dihedral angles are randomly distributed. For each geometry, $\Delta \mathrm{E_{ST}}$ has been calculated at the TDDFT level of theory using the w-B2GPLYP functional. The results obtained with Wigner-sampled geometries at 300 K is presented in FIG.~\ref{fig:box-whisker-all} and the corresponding results at 0 K is given in  FIG. S2 of the supplementary material.

The box and whisker plots indicate that while molecule \textbf{1} gains better average stabilization due to double excitation, the average stabilization due to exchange interactions favors molecule \textbf{2}. However, the combination of these two factors reveal that the average $\Delta \mathrm{E_{ST}}$ value of molecule \textbf{1} and molecule \textbf{2} will be -0.05 eV and +0.03 eV, indicating that molecule \textbf{1} will retain its INVEST property in the Wigner sampled geometries (see FIG.~\ref{fig:box-whisker-all}). It is also worth noting that $\Delta \mathrm{E_{ST}}$ for both molecules  may have positive and negative values, respectively, as is evident from the scattered data as seen in FIGs. S3 and S4 of the supplementary material. This also suggests that the INVEST phenomenon observed in molecule \textbf{1} is not of purely electronic origin, and that the effects of molecular vibrations need to be taken into account for a complete understanding the dynamics of the phenomenon.  

After inspecting the $\Delta \mathrm{E_{ST}}$ values from the above box and whisker plots, we have monitored how the average $\Delta \mathrm{E_{ST}}$ values are changing with increase and decrease of all bond lengths, bond angles  and dihedral angles involved with the heptazine moiety with respect to it equilibrium configuration and prepared a large set of box and whisker plots. We have identified three dihedral angles, namely  $\angle \mathrm{C_{3}-N_{2}-C_{1}-N_{4}}$, $\angle \mathrm{C_{1}-N_{4}-C_{5}-N_{6}}$ and $\mathrm{N_{2}-C_{1}-N_{4}-C_{5}}$, where a change in the dihedral angle shifts the mean and median of the box and whisker plot (see FIG.~\ref{fig:box-whisker-plot-three-selected-da-300K}), indicating that the INVEST property of molecule \textbf{1} is sensitive to the Wigner sampled geometries obtained in the harmonic region. Similar characteristic box and whisker plots of these three dihedral angles are also obtained at 0 K (see FIG. S5 of supplementary material) and the corresponding results for the non-INVEST molecule (\textbf{2}) are provided in FIGs. S6 and S7 of the supplementary material.

We note that a change in the dihedral angle, $\angle \mathrm{C_{1}-N_{4}-C_{5}-N_{6}}$ from the equilibrium geometry leads to the out-of-plane configuration of the central nitrogen (N$_{6}$) and C$_{1}$, while the corresponding changes in $\angle \mathrm{C_{3}-N_{2}-C_{1}-N_{4}}$ and $\mathrm{N_{2}-C_{1}-N_{4}-C_{5}}$ bring  (N$_{4}$)/(C$_{3}$) and (N$_{2}$)/(C$_{5}$) out of the plane. To understand the role of the out of plane motions of the above stated atoms, we have replaced N$_{2}$, N$_{4}$ and N$_{6}$ by phosphorous atoms and found that phosphorous-containing heptazine moieties are nonplanar at their equilibrium geometries. Interestingly, when N$_{6}$ is replaced by phosphorous, the equilibrium dihedral angle  $\angle \mathrm{C_{1}-N_{4}-C_{5}-P_{6}}$ is increased by 14$^\circ$ with respect to $\angle \mathrm{C_{1}-N_{4}-C_{5}-N_{6}}$ of the pristine molecule \textbf{1} and $\Delta \mathrm{E_{ST}}$ of the  P$_{6}$ substituted molecule \textbf{1} is -0.26 eV, which is significantly larger than the calculated $\Delta \mathrm{E_{ST}}$ value (-0.089 eV) of the unsubstituted molecule \textbf{1}. 

The combined effect of the substitutions of N$_{2}$, N$_{4}$ by P atoms give a $\Delta \mathrm{E_{ST}}$ value of -0.14 eV, and in this case P$_{2}$ is more out of the plane than P$_{4}$. Both box and whisker plots and the effect of the  phosphorous substitution suggest INVEST is connected to the out of the plane motion of N$_{2}$, N$_{4}$ and N$_{6}$, and this is further substantiated by the fact that if we take the value of $\angle \mathrm{C_{1}-N_{4}-C_{5}-N_{6}}$ of molecule \textbf{2} as that of molecule \textbf{1}, the $\Delta \mathrm{E_{ST}}$ value of molecule \textbf{2} also  becomes negative (-0.02 eV). Furthermore, the replacement of N$_{6}$ of molecule \textbf{2} by a phosphorous atom, molecule \textbf{2} will theoretically exhibit INVEST with a corresponding $\Delta \mathrm{E_{ST}}$ value of -0.49 eV.  It is to be noted that the variation of the inverted energy gap with the change in dihedral angles involving the other nitrogen atoms is small. These results are collected in FIG. S8 of the supplementary file. \\

\begin{figure}
\includegraphics{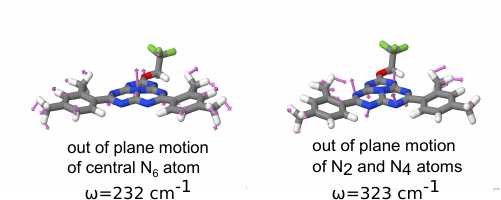}
\caption{\label{fig:normal-modes-motion}Nature of the normal mode motion associated with the movement of the N$_{6}$, N$_{4}$ and N$_{2}$ atoms.}
\end{figure}

Since three particular dihedral angles are important in determining INVEST in molecule \textbf{1}, we need to probe whether or not any of the normal mode motions are involved along these dihedral angles. The normal mode analyses identify an out-of-plane motion of N$_{6}$ with large displacements at 232 cm$^{-1}$, whereas similar motions of N$_{2}$ and N$_{4}$ could be seen in a normal mode having a frequency of 323 cm$^{-1}$ (see FIG.~\ref{fig:normal-modes-motion}). %It is important to mention here that the study on the geometric effects in INVEST wouldn't have been scientifically meaningful without the existence of these normal mode frequencies. 
Based on these findings, we have investigated the variations of the rate constant of rISC with these three dihedral angles at different temperatures using our code. Each of the dihedral angles has been changed up to 10 degrees with a regular interval of 1 degree and k$_{\mathrm{rISC}}$ is computed on this new set of 10 geometries at five different temperatures, namely, 200, 225, 250, 275 and 300 K. 
\begin{figure}
\includegraphics{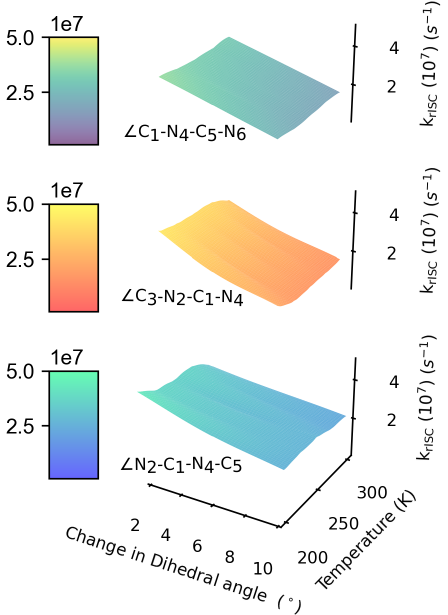}
\caption{\label{fig:krisc-vs-temp-and-da-three}Variation of k$_{\mathrm{rISC}}$ with the change in dihedral angle at different temperatures (within the range of 200-300 K) of three selected dihedral angles.}
\end{figure}

The relevant results for all three dihedral angles, namely, $\angle \mathrm{C_{1}-N_{4}-C_{5}-N_{6}}$, $\angle \mathrm{C_{3}-N_{2}-C_{1}-N_{4}}$ and $\angle \mathrm{N_{2}-C_{1}-N_{4}-C_{5}}$ are depicted in FIG.~\ref{fig:krisc-vs-temp-and-da-three}. The surfaces generated from these 3D plots manifestly show that k$_{\mathrm{rISC}}$ varies only slightly with temperature regardless of the change of the dihedral angles from their equilibrium values, and these findings suggest a possible microscopic origin for the slow change in the rate constant of rISC with temperature in the INVEST molecule. \\

 To conclude, we have calculated the temperature-dependent rate constant of rISC for a pair of molecules featuring inverted and non-inverted singlet-triplet gap. To understand the origin of the inverted singlet-triplet gap (INVEST), we have computed the exchange energy and doubles-excitation contributions on a set of geometries generated from a Wigner phase space sampling. Our study reveals that the out-of-plane motion of nitrogen atoms in three specific dihedral angles of the heptazine moiety are instrumental in determining the INVEST property. Furthermore, the k$_{\mathrm{rISC}}$ values obtained with geometries created when changing these dihedral angles independently in the temperature range 200-300 K, demonstrate that the change of these dihedral angles during normal mode vibrations alter the values of the
rate constant only to a limited extent and this acts as a possible microscopic reason for the origin of the experimentally observed temperature dependent delayed fluorescence lifetime pattern of the INVEST molecule. In brief, the present work provides substantial evidence in favor of the vibronic origin of INVEST and concomitant temperature-dependent reverse intersystem crossing process occurring in the INVEST molecule.  

\begin{acknowledgments}
P. K. thanks the Council of Scientific and Industrial Research (CSIR) for granting him the Senior Research Fellowship. K.R. has received support from the Research Council of Norway through a Centre of Excellence Grant (grant number 262695). 
 The support and resources provided by ‘PARAM Shakti Facility’ under the National Supercomputing Mission, Government of India at the Indian Institute of Technology, Kharagpur, are gratefully acknowledged by S.C.
\end{acknowledgments}

\bibliography{manuscript}% Produces the bibliography via BibTeX.

\end{document}

\preprint{APS/123-QED}

\title{Supplementary Material \\
of \\
Reverse intersystem crossing dynamics in Vibronically modulated inverted singlet-triplet gap system: A Wigner phase space study }% Force line breaks with \\
\blfootnote{A footnote to the article title}%

\author{Pijush Karak}
\altaffiliation{Department of Chemistry, University of Calcutta, 92 A. P. C. Road,
Kolkata–700009, West Bengal, India.}
\author{Pradipta Manna}
\altaffiliation{Department of Chemistry, University of Calcutta, 92 A. P. C. Road,
Kolkata–700009, West Bengal, India.}
\author{Ambar Banerjee$^{*}$}
\altaffiliation{Research Institute for Sustainable Energy (RISE), TCG Centres for Research and Education in Science and Technology (TCG-CREST), Kolkata, 700091 India
}
\email{ambar.banerjee@tcgcrest.org.}
\author{Kenneth Ruud$^{*}$}
\altaffiliation{Hylleraas Centre for Quantum Molecular Sciences, Department of Chemistry,
University of Tromsø – The Arctic University of Norway, 9037 Tromsø, Norway}
\altaffiliation{Norwegian Defence Research Establishment, P.O.Box 25, 2027 Kjeller, Norway}
\email{kenneth.ruud@uit.no.}

\author{Swapan Chakrabarti$^{*}$}
\altaffiliation{Department of Chemistry, University of Calcutta, 92 A. P. C. Road, Kolkata–700009, West Bengal, India}
\email{swcchem@caluniv.ac.in.}


\date{\today}% It is always \today, today,

\maketitle

%\tableofcontents
\tableofcontents
\newpage
\section{Theoretical Method}
 \subsection{Rate of Intersystem crossing (ISC) and reverse intersystem crossing (rISC) driven by direct-spin-orbit (DSO) interaction}
The total Hamiltonian with DSO interaction can be written as,
\begin{equation}
  \hat{H}^{\prime}=\hat{H}_{\mathrm{0}}+\hat{H}_{\mathrm{SO}},  
  \label{eqn:total-hamiltonian}
\end{equation}
where, $\hat{H}_{\mathrm{0}}$ and $\hat{H}_{\mathrm{SO}}$ are the unperturbed and DSO Hamiltonians, respectively.

Employing the first-order perturbation theory, the Fermi's Golden rule based expressions for $\mathrm{k_{ISC}^{\mathrm{DSO}}}$ between S$_{\mathrm{1}}$ and the T$_{\mathrm{1}}$  and $\mathrm{k_{rISC}}$ between T$_{\mathrm{1}}$ and  S$_{\mathrm{1}}$ can be written as

\begin{align}
  \mathrm{k_{ISC}^{\mathrm{DSO}}}&=\frac{2\pi}{\hbar z}\sum_{v_{\mathrm{S_{1}}},v_{\mathrm{T_{1}}}}e^{-\beta E_{v_{\mathrm{S_{1}}}}} 
  \lvert \langle {\mathrm{S_{1}},v_{\mathrm{S_{1}}} \lvert \hat{H}_{\mathrm{SO}} \rvert \mathrm{T_{1}},v_{\mathrm{T_{1}}}} \rangle \rvert ^{2} 
  \times \delta (\Delta E _{\mathrm{ST}}+E_{v_{\mathrm{S_{1}}}}-E_{v_{\mathrm{T_{1}}}})~.
 \label{eqn:kisc-dso-part}
 \end{align} and 
\begin{align}
  \mathrm{k_{rISC}^{\mathrm{DSO}}}&=\frac{2\pi}{\hbar z}\sum_{v_{\mathrm{T_{1}}},v_{\mathrm{S_{1}}}}e^{-\beta E_{v_{\mathrm{T_{1}}}}} 
  \lvert \langle {\mathrm{T_{1}},v_{\mathrm{T_{1}}} \lvert \hat{H}_{\mathrm{SO}} \rvert \mathrm{S_{1}},v_{\mathrm{S_{1}}}} \rangle \rvert ^{2} 
  \times \delta (-\Delta E _{\mathrm{ST}}-E_{v_{\mathrm{S_{1}}}}+E_{v_{\mathrm{T_{1}}}})~,
 \label{eqn:krisc-dso-part}
\end{align}
respectively. 

The energy of the electronic (singlet and triplets) and vibrational states  in  Eqs~\ref{eqn:total-hamiltonian}, \ref{eqn:kisc-dso-part} and \ref{eqn:krisc-dso-part}, are denoted by E$_{\mathrm{S_{1}/T_{1}}}$ and E$_{v_{\mathrm{S_{1}/T_{1}}}}$. The singlet-triplet energy gap is defined by $\Delta E_{\mathrm{ST}}$. $z$ is the canonical partition function of the initial electronic state. The expression of $\beta$ is $\frac{1}{kT}$, where $k$ and $T$ are the Boltzmann constant and temperature, respectively. 
The final expressions of k$_{\mathrm{ISC}}^{\mathrm{DSO}}$ and k$_{\mathrm{rISC}}^{\mathrm{DSO}}$ equations (Eqs.~\ref{eqn:kisc-dso-part} and \ref{eqn:krisc-dso-part}), in terms of generating functions can be expressed as,
\begin{align}
k_{\mathrm{ISC}}^{\mathrm{DSO}}= \frac{\vert {{H}}_{\mathrm{SO}}\vert^{2}}{\hbar^{2} z}\int_{-\infty}^{\infty}G_{\mathrm{DSO}}^{\mathrm{ISC}}(t,t^\prime)e^{i\Delta E_{\mathrm{ST}}\frac{t}{\hbar}} dt,
\label{eqn:kisc-gdso-part}
\end{align}
and 
\begin{align}
k_{\mathrm{rISC}}^{\mathrm{DSO}}= \frac{\vert {{H}}_{\mathrm{SO}}\vert^{2}}{\hbar^{2} z}\int_{-\infty}^{\infty}G_{\mathrm{DSO}}^{\mathrm{rISC}}(t,t^\prime)e^{-i\Delta E_{\mathrm{ST}}\frac{t}{\hbar}} dt,
\label{eqn:krisc-gdso-part}
\end{align}
respectively.
The form of the ${H}_{\mathrm{SO}}$ is 
\begin{align}
    {H}_{\mathrm{SO}}=\langle {\mathrm{S_{1}} \lvert \hat{H}_{\mathrm{SO}} \rvert \mathrm{T_{1}}} \rangle .
\end{align}

In Eqs.~\ref{eqn:kisc-gdso-part} and ~\ref{eqn:krisc-gdso-part}, $G_{\mathrm{DSO}}^{\mathrm{ISC}}(t,t^\prime)$ and $G_{\mathrm{DSO}}^{\mathrm{rISC}}(t,t^\prime)$ represent the time-dependent correlation functions obtained with the direct spin-orbit (DSO) of ISC and rISC mechanisms, respectively and $t^\prime = -\frac{t}{\hbar}-i\beta$ where t indicates the time. 

The general expression of correlation function ($G_{\mathrm{DSO}}^{\mathrm{ISC}}(t,t^\prime)$/ $G_{\mathrm{DSO}}^{\mathrm{rISC}}(t,t^\prime)$) for ISC/rISC is written as, 
\begin{align}
G_{\mathrm{DSO}}^{\mathrm{ISC/rISC}}(t,t^\prime)&=\sqrt{\frac{\mathrm{det}({\bf{S}}_{i}){\mathrm{det}}({\bf{S}}_{f})}{\mathrm{det}({\bf{W}})}}exp(-\frac{i}{2\hbar}{{\bf{V}}^{\mathrm{T}}{\bf{W}}^{-1}{\bf{V}}}+\frac{i}{\hbar}{\bf{D}}^{\mathrm{T}}{\bf{UD}}).
\label{eq:G-ISC-rISC}
\end{align}
The detailed derivation of the above correlation functions (Eq.~\ref{eq:G-ISC-rISC}) can be found in Refs~\cite{Peng_jctc_2013_correlation_function_triplet_excited_decay_soc_nonadiabatic_coupling,Kim_jctc_2020_spin_vibronic_risc_tadf_solvent,Pijush_jcp_2022_spin_vibronic_risc,KARAK2023}.
In the above Eq.~\ref{eq:G-ISC-rISC}, the form of ${\bf{S}}_{f}$, $ {\bf{B}}_{f}$, ${\bf{S}}_{i}$, ${\bf{B}}_{i}$, $\bf{W}$, $\bf{U}$ and $\bf{V}$ are,
\begin{eqnarray}
    {\bf{S}}_{f}=\frac{\omega_{f}}{sin(\omega_{f}t\hbar)}, ~~\\
     {\bf{B}}_{f}=\frac{\omega_{f}}{tan(\omega_{f}t\hbar)} , ~~\\   {\bf{S}}_{i}=\frac{\omega_{i}}{sin(\omega_{i}t^\prime\hbar)}, ~~\\
     {\bf{B}}_{i}=\frac{\omega_{i}}{tan(\omega_{i}t^\prime\hbar)}, ~~\\ 
    \\
   %\end{eqnarray}
%as well as,
%   \begin{eqnarray}
    {\bf{W}}=\left [
    \begin{array}{cc}
        {\bf{B}}_{f}+{\bf{J}}^{\mathrm{T}}{\bf{B}}_{i}{\bf{J}} & -({\bf{S}}_{f}+{\bf{J}}^{\mathrm{T}}{\bf{S}}_{i}{\bf{J}}) \\
        -({\bf{S}}_{f}+{\bf{J}}^{\mathrm{T}}{\bf{S}}_{i}{\bf{J}}) &  {\bf{B}}_{f}+{\bf{J}}^{\mathrm{T}}{\bf{B}}_{i}{\bf{J}}
    \end{array}
    \right ], \\
    {\bf{U}}=({\bf{B}}_{i}-{\bf{S}}_{i}) ~~ \\    
    \textrm{and} \nonumber \\
    {\bf{V}}=\left [
    \begin{array}{c}
         {\bf{J}}^{\mathrm{T}}{\bf{UD}} \\
         {\bf{J}}^{\mathrm{T}}{\bf{UD}}
    \end{array}
    \right ],~~ 
    \end{eqnarray}
    \\
 respectively.   
   From Eqs.~S7-S15, the initial ($i$) and final ($f$) states for ISC (S$_{1}\rightsquigarrow$T$_{1}$) and rISC (T$_{1}\rightsquigarrow$S$_{1}$) are S$_{1}$ and T$_{1}$ and T$_{1}$ and S$_{1}$, respectively. The frequency, displacement vector and Duschinsky rotation matrix are denoted by $\omega, {\bf{D}}$ and $\bf{J}$, respectively. Duschinsky rotation matrix connects the normal coordinates of the initial (${\bf{Q}}_{i}$) and final (${\bf{Q}}_{f}$) electronic states by the relation ${\bf{Q}}_{f}={\bf{JQ}}_{i}+{\bf{D}}$. 
   ${\bf{S}}$ and ${\bf{B}}$ are diagonal matrices. $\bf{W}$ and $\bf{X}$ are ($2N \times 2N$) matrices. $\bf{V}$, ${\bf{Y}}$ and {\bf{U}} are ($2N \times 1$), ($2N \times 1$) and ($N \times 1$) column vectors, respectively.

\section{Computational Details}
\begin{figure*}[htb!] 
\centering
\includegraphics[width=\textwidth]{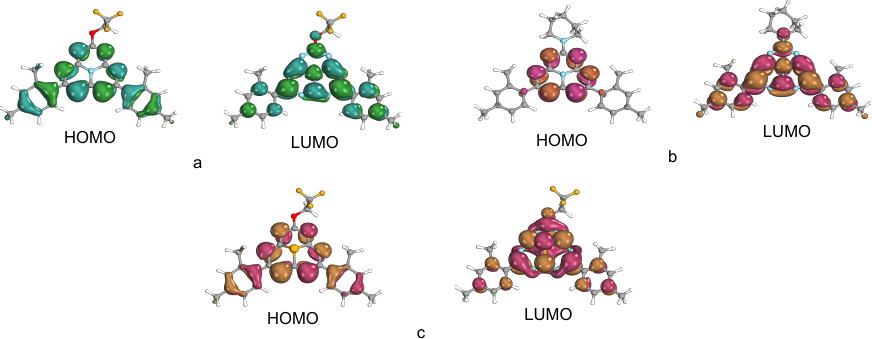}
    \caption{HOMO and LUMO of (a) \textbf{1} (b) \textbf{2} and (c) phosphorus substituted (replacing central N atom by P atom) molecules obtained using $\omega$B2GPLYP functional.}
    \label{fig:homo-lumo-wb2gplyp}
\end{figure*}

The geometry of the ground state singlet (S$_{0}$) and triplet (T$_{1}$)  were optimized at the level of density functional theory (DFT) using B3LYP functional and 6-311G(d,p) basis set. Time-dependent DFT has been used for the optimization of the S$_{1}$ state using the same functional and basis sets. The absence of any imaginary frequency in the S$_{0}$, T$_{1}$ and S$_{1}$ geometries confirm that all structures are located to their respective global minima of the potential energy surfaces. The geometry optimizations and frequency calculations were performed in Gaussian16 software.
The transition orbitals associated with the lowest excited singlet (S$_{1}$) and the triplet (T$_{1}$) states are highest occupied molecular orbital (HOMO) and lowest unoccupied molecular orbitals (LUMO) as depicted in Figure.~\ref{fig:homo-lumo-wb2gplyp}. 
The HOMO and LUMO computed using the doubles corrected $\omega$B2GPLYP hybrid functional and cc-pVDZ basis set at the equilibrium geometry of S$_{1}$ indicate that the overlap between these two orbitals is small as the electron density in these two orbitals is located on different atoms. The decrease in overlap thereby hinders the exchange interaction between them and consequently the T$_{1}$ state is less stabilized. The computed value of the exchange integral among HOMO and LUMO at the equilibrium geometry of S$_{1}$ of INVEST (\textbf{1}) molecule is only 0.15 eV and  after taking the doubles-excitation contribution, the S$_{1}$ state is more stabilized compared to that of the T$_{1}$ state. This is because of the larger configurations arising due to doubles excitation in the singlet state whereas the Pauli exclusion principle precludes such possibility for the triplet electronic state. Therefore, the extra stability gained by the S$_{1}$ state leads to the lowering of energy of this state over the T$_{1}$ state, making an inverted singlet-triplet gap. The calculated inverted energy gap of INVEST molecule (\textbf{1}) using doubles corrected hybrid functional namely, $\omega$-B2GPLYP is -0.089 eV whereas the experimental energy gap is -0.011 eV.  By using the same functional, the computed energy gap between S$_{1}$ and T$_{1}$ of non-INVEST molecule (\textbf{2}) has been found to be +0.01 eV. The experimentally observed energy gap of molecule \textbf{2}  was found to be +0.052 eV. The computation of the singlet-triplet energy gap was done using  ORCA 502. To understand the role of vibrations on the singlet-triplet energy gap, we have performed Wigner phase space sampling using the equilibrium geometry of the S$_{1}$ state both for INVEST and non-INVEST molecules. Using the SHARC program package,\cite{Mai_jcp_2017_isc_sharc,Mai_2018_WIRES_SHARC_nonadiabatic_dynamics} we have generated 300 Wigner sampled geometries both at 0 K and 300 K. On each geometry of the generated ensemble at 0 K and 300 K, a single point calculation was done considering 5 lowest excited singlet and triplet states using  $\omega$-B2GPLYP, in combination with cc-pVDZ basis set and its auxiliary part, namely cc-pVDZ/C. Additionally, to accelerate the two electron integrals, RIJCOSX approximation has been used.  At finite temperature (300 K), we have actually made an ensemble of 600 geometries and then on each geometry, TDDFT single point calculation was performed. Unfortunately, we have found that the computations of the excitation energies of the triplet states on some geometries were forbidden due to convergence failure and therefore, we have rejected those geometries for further data collections and but overall we have found 322 geometries that correctly prints the eigenvectors of the triplet state; of which we have considered first 300 geometries for the computation of the triplet state excitation energies. The doubles correction and the exchange interaction energy calculations were carried out on each geometry of the generated ensemble at 0 K and 300 K for both INVEST and non-INVEST molecules.  

\section{ISC and rISC rate constants of the studied systems between the lowest excited singlet and the triplet states}
\begin{table*}[htb!]
\centering
  \caption{Energy gap, SOCME, ISC and  rISC rate constant between the S$_{1}$ and the T$_{1}$ states. Here a and b denote the rate constants obtained by theory and experiments, respectively.}
  \begin{tabular}{lllllll}
    \hline
    System & $\Delta \mathrm{E_{ST}}$ (eV)  & SOCME (cm$^{-1}$)& k$_{\mathrm{rISC}}^{a}$ (s$^{-1}$) & k$_{\mathrm{rISC}}^{b}$ (s$^{-1}$)& k$_{\mathrm{ISC}}^{a}$ (s$^{-1}$) & k$_{\mathrm{ISC}}^{b}$ (s$^{-1}$)\\
    \hline
    \textbf{1} & -0.089 & 1.01 & 2.6 $\times 10^{7}$ & 4.2 $\times 10^{7}$ & 0.7 $\times 10^{7}$& 2.3 $\times 10^{7}$ \\ 
    \textbf{2} & 0.01 & 0.80 & 1.0$\times 10^{7}$ & 2.2 $\times 10^{7}$ &1.5 $\times 10^{7}$ &8.9 $\times 10^{7}$\\
    \hline   
   \label{tbl:kisc-krisc-equilibrium-geom}
  \end{tabular}
\end{table*} 
  CASSCF/NEVPT2 (6e,6o) method was used to compute the SOCME using the cc-pVDZ basis set at the equilibrium geometry of the S$_{1}$ state in ORCA 502 software.\\

\clearpage

\section{Effect of substitution of nitrogen atom by phosphorus on the singlet-triplet energy gap}
To understand the role of the central N atom (N$_{6}$) and other peripheral N atoms such as N$_{2}$ and N$_{4}$ on the energy gap between S$_{1}$ and T$_{1}$ states, we have replaced N$_{6}$, N$_{2}$, N$_{4}$ and N$_{2}$N$_{4}$ atoms of INVEST molecule by P atom individually, producing 4 distinct phosphorus substituted molecules. These 4 hypothetical molecules were also optimized at the same level of theory that has been used for INVEST and non-INVEST molecules as discussed in the computational details section. The change in values of three selected dihedral angles and the singlet-triplet energy gaps of INVEST (\textbf{1}), non-INVEST (\textbf{2}) and 4 phosphorus substituted molecules are provided in TABLE.~\ref{tbl:geometrical-parameters-energy-gap}. The magnitude of these selected  dihedral angles are largely deviated in four P substituted molecules compared to the INVEST and non-INVEST molecules  with a significant change in the energy gap between the S$_{1}$ and T$_{1}$ states (see TABLE.~\ref{tbl:geometrical-parameters-energy-gap}) in (P$_{6}$) and P$_{2}$P$_{4}$ substituted molecules. This observation indicates that the vibrational motion associated with these nitrogen atoms may have deep impact on the singlet-triplet energy gap of the INVEST molecule.\\
\begin{table*}[htb!]
\caption{\label{tbl:geometrical-parameters-energy-gap}Energy gap and the values of three selected dihedral angles of \textbf{1} (INVEST), \textbf{2} (non-INVEST) and phosphorus substituted molecules, computed at the equilibrium geometry of S$_{1}$. Here P(Nn) indicates the substitution of N by a P atom where n could be 2,4 and 6.}
 \begin{tabular}{|l|lll|l|}
    \hline 
   \multicolumn{1}{|c|}{Molecule} &
      \multicolumn{3}{c|}{Dihedral angle } &
      \multicolumn{1}{c|}{Energy gap} \\
      \hline
       & $\angle$ C$_{1}$-N$_{4}$-C$_{5}$-N$_{6}$ ($^\circ$) & $\angle$ C$_{3}$-N$_{2}$-C$_{1}$-N$_{4}$ ($^\circ$) & $\angle$ N$_{2}$-C$_{1}$-N$_{4}$-C$_{5}$ ($^\circ$) & S$_{1}$-T$_{1}$ (eV)\\ \hline
\textbf{1} & 2.82  & -12.42 & 12.80 &-0.089 \\ \hline
\textbf{1}-P(N6) & 16.81 & -23.62 & 25.12 &-0.26  \\ \hline 
\textbf{1}-P(N4) &  0.35 &  0.40 & -0.84 &0.15  \\ \hline 
\textbf{1}-P(N2) & -8.29 & 12.33 & -6.95 &-0.003  \\ \hline
\textbf{1}-P(N2)P(N4) & -4.56 & 6.95 & -2.32 &-0.14  \\ \hline
\textbf{2} & 1.36 & -15.23 & 15.60 & 0.01 \\ \hline
  \end{tabular}
\end{table*}
We have also computed the ISC and rISC rate constantS for the molecule having central nitrogen atom (N$_{6}$) of INVEST molecule  replaced by the P atom. Using the computed energy gap and SOC, the calculated values of k$_{\mathrm{ISC}}$ and k$_{\mathrm{rISC}}$ are found to be 1.16 $\times 10^{7}$  and  1.28 $\times 10^{6}$ s$^{-1}$, respectively. The results are presented in TABLE.~\ref{tbl:kisc-krisc-equilibrium-geom-p-compound}. The larger K$_{\mathrm{rISC}}$ over  k$_{\mathrm{ISC}}$ demonstrates the presence of the inverted energy gap ($\Delta {\mathrm{E_{ST}}}$) in this P substituted molecule. 
\begin{table*}[htb!]
\centering
  \caption{Energy gap, SOCME, ISC and rISC rate constants between the S$_{1}$ and the T$_{1}$ states of P substituted molecule.}
  \begin{tabular}{llll}
    \hline
    $\Delta \mathrm{E_{ST}}$ (eV)  & SOCME (cm$^{-1}$)& k$_{\mathrm{rISC}}$ (s$^{-1}$) (theor.) & k$_{\mathrm{ISC}}$ (s$^{-1}$) (theor.) \\
    \hline
    -0.26 & 0.60 & 1.16 $\times 10^{7}$ & 1.28 $\times 10^{6}$  \\ 
     \hline
   \label{tbl:kisc-krisc-equilibrium-geom-p-compound}
  \end{tabular}
\end{table*} 

\clearpage

\section{Box and Whisker plot of distribution of doubles correction, exchange energy and energy gap at 0 K of INVEST and non-INVEST molecules }

\begin{figure*}[htb!]
\includegraphics{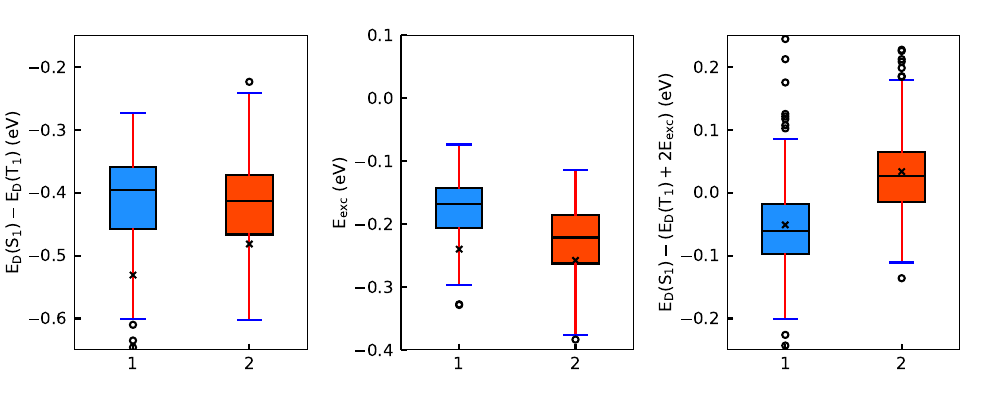}
\caption{\label{fig:box-whisker-all-0K}Box-whisker plot for representing the contribution of the doubles correction to the S$_{1}$ and the T$_{1}$ state (left), exchange energy (middle) and the net effect of doubles and exchange (right) computed on Wigner phase space sampled 300 geometries at 0 K of molecules \textbf{1} and \textbf{2}. The median and mean values are denoted by black solid line and cross point, respectively.  }
\end{figure*}

The computed doubles correction, exchange energy and singlet-triplet energy gap ($\Delta \mathrm{E_{ST}}$) on Wigner sampled geometries at 0 K is given in FIG.~\ref{fig:box-whisker-all-0K}. In this figure, the doubles corrected energy for the singlet and the triplet states are denoted by E$_{\mathrm{D}}(\mathrm{S_{1}})$ and E$_{\mathrm{D}}(\mathrm{T_{1}})$, respectively. Here exchange energy is designated as E$_{\mathrm{exc}}$ and the $\Delta \mathrm{E_{ST}}$ value is computed using the following relation
\begin{equation}
   \Delta \mathrm{E_{ST}} = \mathrm{E_{D}(S_{1})-(E_{D}(T_{1})+2E_{exc})}.~~
\end{equation}
FIG.~\ref{fig:box-whisker-all-0K} implies that at 0 K, the average stability gain of the S$_{1}$ state of the molecule \textbf{1} is more compared to that of the T$_{1}$ state. This happens due to the fact that the stabilization of the T$_{1}$ state through  exchange interaction (see FIG.~\ref{fig:box-whisker-all-0K} (middle panel)) is more than compensated by the larger doubles contribution associated with the S$_{1}$ state (see FIG.~\ref{fig:box-whisker-all-0K} (left panel)). As a result, the net effect of these two factors gives rise to the inversion of the S$_{1}$ and the T$_{1}$ states (see FIG.~\ref{fig:box-whisker-all-0K} (right panel)) in molecule \textbf{1}. On the contrary, the reverse trend has been found in case of molecule \textbf{2}.

\clearpage

\section{Distribution of the computed energy gap  on the Wigner sampled geometries at 0 and 300 K for INVEST and non-INVEST molecules through scatter plot}
 \begin{figure*}[htb!]
\includegraphics{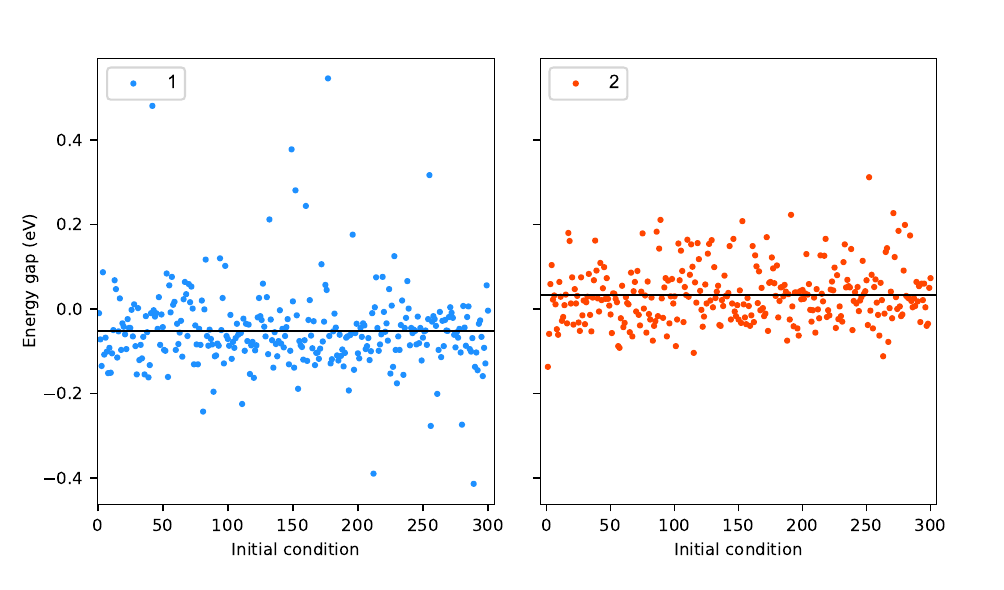}
\caption{\label{fig:energy-gap-wigner-sampling-0K}Distribution of the singlet-triplet energy gap in eV on Wigner sampling geometries at 0 K of INVEST (\textbf{1}) (left) and non-INVEST (\textbf{2}) (right) molecules. The black line indicates the mean value of the energy gap computed over the Wigner sampling geometries.}
\end{figure*}

 \begin{figure*}[htb!]
\includegraphics{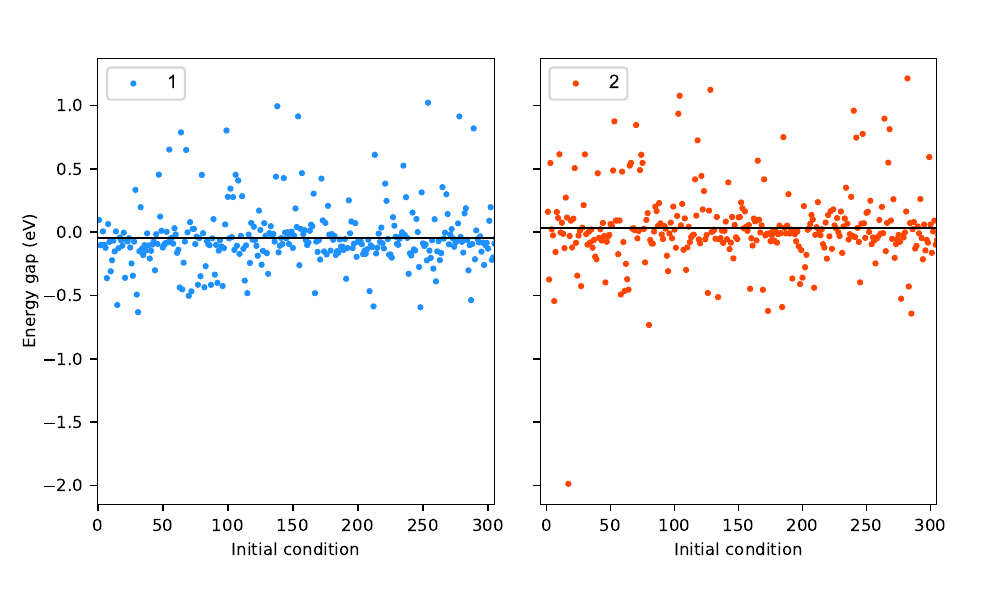}
\caption{\label{fig:energy-gap-wigner-sampling-300K}Distribution of the computed singlet-triplet energy gap in eV on Wigner sampling geometries at 300 K of INVEST (\textbf{1}) (left) and non-INVEST (\textbf{2}) (right) molecules. The black line indicates the mean value of the energy gap calculated over the Wigner sampling geometries. }
\end{figure*}

From FIGs.~\ref{fig:energy-gap-wigner-sampling-0K}
 and ~\ref{fig:energy-gap-wigner-sampling-300K}, it is evident that in molecule \textbf{1} and \textbf{2}, the mean values (black solid line) are located in the negative and positive position of the energy gap axis, indicating that molecules \textbf{1} and \textbf{2} exhibit INVEST and non-INVEST behaviors, respectively. 
 It is worth noting that the more dispersed data at 300 K as shown in FIG.~\ref{fig:energy-gap-wigner-sampling-300K} is an indication of the Boltzmann distribution at finite temperature in Wigner phase space sampling. \\

\clearpage
 
 \section{Representation of the change in energy gap with the change in dihedral angle of three selected dihedral angles through Box and Whisker plots}
\begin{figure*}[htb!]
\includegraphics{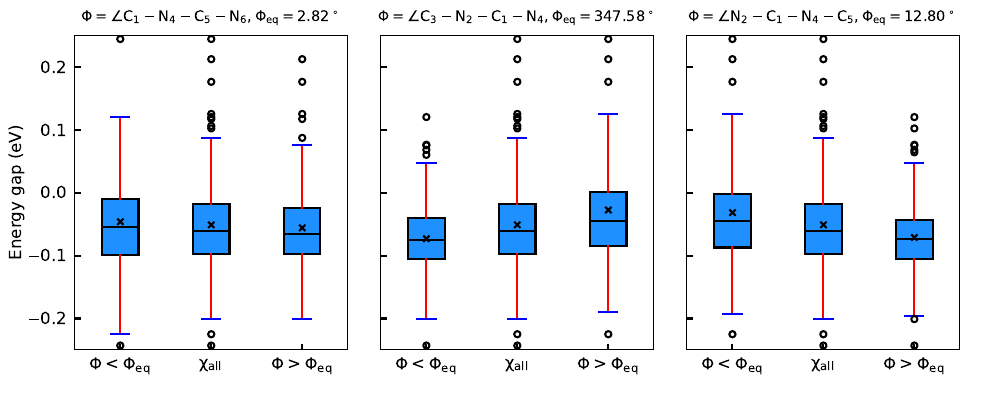}
\caption{\label{fig:box-whisker-plot-three-selected-da-0K} Distribution of the singlet-triplet energy gaps computed on Wigner distributed geometries at 0 K of three selected dihedral angles, namely, $\angle \mathrm{C_{1}-N_{4}-C_{5}-N_{6}}$ (left), $\angle \mathrm{C_{3}-N_{2}-C_{1}-N_{4}}$ (middle) and $\angle \mathrm{N_{2}-C_{1}-N_{4}-C_{5}}$ (right) are shown by Box-whisker plots of molecule \textbf{1}. Here, $\mathrm{\chi_{all}}$ denotes the inclusion of all 300 geometries. $\mathrm{\Phi}<\mathrm{\Phi_{eq}}$ and $\mathrm{\Phi}>\mathrm{\Phi_{eq}}$ are suggesting the geometries having dihedral angle less and greater than their corresponding equilibrium values ($\mathrm{\Phi_{eq}}$), respectively. }
\end{figure*}

\begin{figure*}[htb!]
\includegraphics{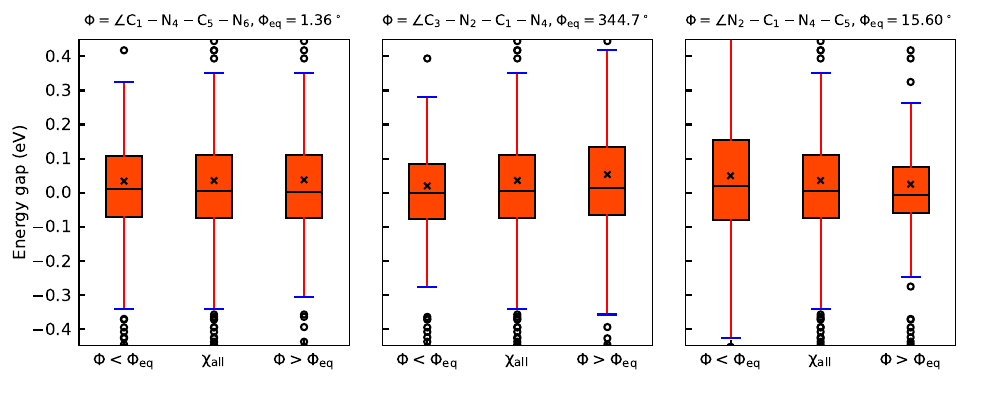}
\caption{\label{fig:box-whisker-plot-three-selected-da-0K}The computed energy gaps between the S$_{1}$ and the T$_{1}$ states of an ensemble containing Wigner distributed geometries at 300 K of three selected dihedral angles, namely, $\angle \mathrm{C_{1}-N_{4}-C_{5}-N_{6}}$ (left), $\angle \mathrm{C_{3}-N_{2}-C_{1}-N_{4}}$ (middle) and $\angle \mathrm{N_{2}-C_{1}-N_{4}-C_{5}}$ (right) are depicted by Box-whisker plots of molecule \textbf{2} .}
\end{figure*}
\clearpage

\begin{figure*}[htb!]
\includegraphics{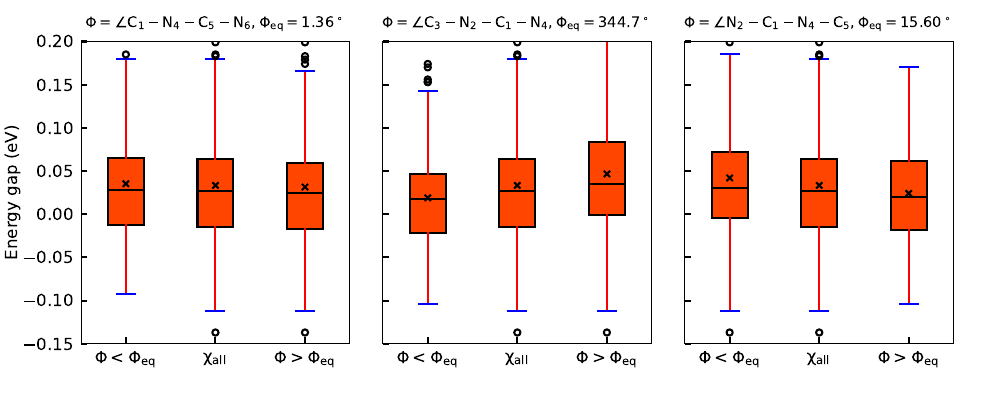}
\caption{\label{fig:box-whisker-plot-three-selected-da-300K-HZPIPX2}Box-whisker plot of representing the distribution of the  singlet-triplet energy gap calculated based on the Wigner phase space sampling at 0 K of three selected dihedral angles, namely, $\angle \mathrm{C_{1}-N_{4}-C_{5}-N_{6}}$ (left), $\angle \mathrm{C_{3}-N_{2}-C_{1}-N_{4}}$ (middle) and $\angle \mathrm{N_{2}-C_{1}-N_{4}-C_{5}}$ (right) of molecule \textbf{2}.}
\end{figure*}

%\clearpage
\section{Effect of the singlet-triplet energy gap with the variation of dihedral angle associated with the heptazine moiety}

\begin{figure*}[htb!]
\includegraphics{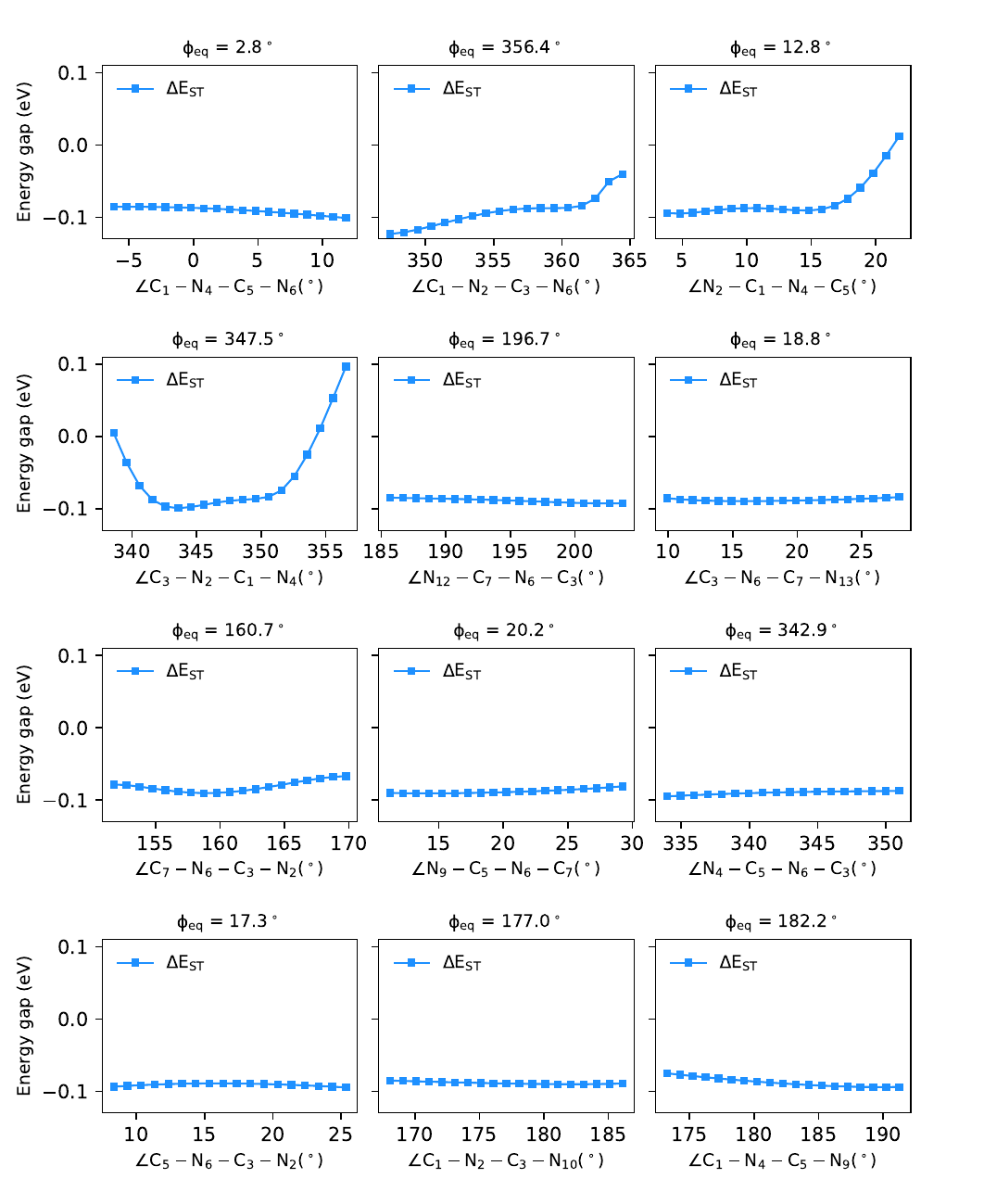}
\caption{\label{fig:energy-gap-vs-dihedral-angle-all}Variation of the singlet-triplet energy gap ($\Delta \mathrm{E_{ST}}$) in eV with the change in dihedral angle of few selected dihedral angles associated with the heptazine part of the INVEST molecule (\textbf{1}). The equilibrium value ($\mathrm{\Phi_{eq}}$) of the dihedral angle is given on top of each subfigure.}
\end{figure*}

We have considered a total of 12 dihedral angles namely, $\angle \mathrm{C_{1}-N_{4}-C_{5}-N_{6}}$, $\angle \mathrm{C_{1}-N_{2}-C_{3}-N_{6}}$, $\angle \mathrm{N_{2}-C_{1}-N_{4}-C_{5}}$, $\angle \mathrm{C_{3}-N_{2}-C_{1}-N_{4}}$, $\angle \mathrm{N_{12}-C_{7}-N_{6}-C_{3}}$, $\angle \mathrm{C_{3}-N_{6}-C_{7}-N_{13}}$, $\angle \mathrm{C_{7}-N_{6}-C_{3}-N_{2}}$, $\angle \mathrm{N_{9}-C_{5}-N_{6}-C_{7}}$, $\angle \mathrm{N_{4}-C_{5}-N_{6}-C_{3}}$, $\angle \mathrm{C_{5}-N_{6}-C_{3}-N_{2}}$, $\angle \mathrm{C_{1}-N_{2}-C_{3}-N_{10}}$ and $\angle \mathrm{C_{1}-N_{4}-C_{5}-N_{9}}$ to understand the effect of the puckering of the heptazine moiety  on the singlet-triplet energy gap. For this purposes, every selected dihedral angle has been changed up to 10 degrees in the positive and negative directions with respect to the equilibrium value with an interval of 1 degree and on each distorted geometry, the energy gap was computed using $\omega$B2GPLYP. The variations of the energy gap with the change in the dihedral angles  are presented in FIG.~\ref{fig:energy-gap-vs-dihedral-angle-all}. From this figure, it is clear that, among the 12 selected dihedral angles, the change of the energy gap has been found to be prominent in those dihedral angles which are associated with the motion of N$_{2}$, N$_{4}$ and N$_{6}$atoms. This finding suggest that the normal mode motion involving these nitrogen atoms can affect the singlet-triplet energy gap of the INVEST molecule.

\clearpage

\section{Coordinates of the optimized geometries of INVEST (\textbf{1}) and non-INVEST (\textbf{2}) molecules}
S$_0$ geometry (\AA) of INVEST molecule (\textbf{1})
\begin{verbatim}
 C                  0.67107000    2.74980400   -0.61580600
 N                  1.62908900    1.83169100   -0.55211000
 C                  1.23735200    0.56844700   -0.37938700
 N                 -0.64589200    2.58800300   -0.53954800
 C                 -1.07273900    1.34120700   -0.36343400
 N                 -0.13185200    0.30129300   -0.27641200
 C                 -0.56960100   -1.02193600   -0.09188700
 C                 -2.72191500   -0.22672700   -0.08789100
 N                 -2.36217600    1.05449900   -0.26734400
 N                  2.10678400   -0.42411900   -0.30531100
 C                  1.63233400   -1.67154600   -0.12380600
 N                 -1.86752500   -1.26381300    0.00088200
 N                  0.33302800   -1.99202400   -0.01606800
 C                  4.00895700   -2.63881600   -0.08412900
 C                  4.78705000   -3.79854800    0.00892800
 C                  4.25183600   -5.07888800    0.14426200
 C                  2.85893900   -5.20443100    0.18868200
 C                  2.05726200   -4.08309800    0.10171500
 C                  2.59675400   -2.78777000   -0.03664800
 H                  5.86632500   -3.69008100   -0.02663500
 H                  2.40446100   -6.18429200    0.29129500
 H                  0.98100100   -4.17764600    0.13708200
 C                 -5.20468600    0.41373700   -0.02715100
 C                 -4.15956600   -0.54806800    0.02152800
 C                 -4.49704800   -1.90747400    0.18336400
 C                 -5.80809900   -2.32782100    0.29384800
 C                 -6.85155900   -1.39662100    0.24938300
 C                 -6.51949800   -0.05190100    0.08938100
 H                 -3.68926500   -2.62495900    0.21786300
 H                 -6.02657500   -3.38370300    0.41479600
 H                 -7.32158100    0.67826600    0.05217200
 C                  5.14168100   -6.29032400    0.25298300
 H                  6.18574900   -6.04121100    0.05592800
 H                  4.83702700   -7.06769600   -0.45354100
 H                  5.08441200   -6.72636800    1.25585400
 C                 -8.28685700   -1.83607100    0.38416800
 H                 -8.51296200   -2.65940900   -0.29937400
 H                 -8.49021200   -2.19471100    1.39871900
 H                 -8.97882800   -1.01879600    0.17385500
 C                 -5.01806500    1.90318800   -0.19336800
 H                 -4.40088300    2.32647500    0.60014800
 H                 -4.50503500    2.14687800   -1.12455800
 H                 -5.99189400    2.39702100   -0.18562000
 C                  4.74821300   -1.32891100   -0.22330500
 H                  5.82440100   -1.51296300   -0.20920300
 H                  4.48953400   -0.81454700   -1.14974400
 H                  4.49855400   -0.63440000    0.57989300
 O                  1.02647300    4.03043300   -0.80052300
 C                  2.80461600    5.00748600    0.50088700
 F                  4.12066800    5.30808700    0.47062600
 F                  2.12958100    6.14377000    0.74623400
 F                  2.59030100    4.17616000    1.53611100
 C                  2.40847600    4.37919100   -0.82479700
 H                  2.53364600    5.12434200   -1.60943500
 H                  3.03576300    3.50881900   -1.00578100
\end{verbatim}
S$_1$ geometry (\AA) of INVEST molecule (\textbf{1})
\begin{verbatim} 
 C                  0.65457500    2.78526400   -0.74459200
 N                  1.60417800    1.85713800   -0.74825300
 C                  1.23340500    0.59929100   -0.49848400
 N                 -0.66923100    2.58645100   -0.74573700
 C                 -1.08319900    1.35152300   -0.48761500
 N                 -0.12698200    0.34823600   -0.17150600
 C                 -0.56124200   -1.00406000   -0.33358700
 C                 -2.74388600   -0.23917000   -0.27025300
 N                 -2.36277400    1.04547400   -0.50893400
 N                  2.09247500   -0.38924000   -0.53681900
 C                  1.65528500   -1.65771400   -0.30760200
 N                 -1.86078000   -1.26107900   -0.37762700
 N                  0.32896700   -1.96801300   -0.39196900
 C                  3.99462000   -2.57209500    0.11199700
 C                  4.77727800   -3.71322100    0.29871800
 C                  4.26583700   -5.01264500    0.28448500
 C                  2.88971100   -5.16780800    0.08109300
 C                  2.08038000   -4.06486500   -0.10747900
 C                  2.59812200   -2.74891600   -0.10621700
 H                  5.84064700   -3.57730800    0.47065900
 H                  2.45264900   -6.16097200    0.06901000
 H                  1.01896000   -4.19310800   -0.26866300
 C                 -5.16147100    0.39892100    0.21379700
 C                 -4.14879000   -0.56662600   -0.04261700
 C                 -4.50779300   -1.93257000   -0.06099300
 C                 -5.80911300   -2.34667000    0.14599700
 C                 -6.82121700   -1.41036800    0.38702700
 C                 -6.46456700   -0.06107700    0.41923700
 H                 -3.73088100   -2.66045800   -0.25075100
 H                 -6.04543900   -3.40536000    0.11932000
 H                 -7.23724400    0.67447100    0.62052800
 C                  5.16595700   -6.20803700    0.46242400
 H                  6.12203800   -5.92709200    0.90859100
 H                  5.37847700   -6.68412400   -0.50151700
 H                  4.70076100   -6.96469400    1.09983900
 C                 -8.25033100   -1.84497800    0.58673400
 H                 -8.71795600   -2.09947700   -0.37100200
 H                 -8.31049800   -2.73343600    1.22096800
 H                 -8.84882300   -1.05614200    1.04652900
 C                 -4.92240800    1.88748100    0.30417900
 H                 -4.14592200    2.13271400    1.03240600
 H                 -4.58728500    2.30453300   -0.64655100
 H                 -5.84310500    2.39153400    0.60483400
 C                  4.68644800   -1.23133100    0.17942300
 H                  5.73539900   -1.37048100    0.44887700
 H                  4.63540600   -0.69936600   -0.77183800
 H                  4.22712900   -0.57225600    0.91966700
 O                  0.99525800    4.08322200   -0.86470100
 C                  2.77572800    4.89349900    0.55367700
 F                  4.09023900    5.19856500    0.55869700
 F                  2.09794000    5.98532900    0.94714700
 F                  2.56691400    3.93374500    1.47455900
 C                  2.37295400    4.44105600   -0.84091000
 H                  2.48928600    5.28375200   -1.52100900
 H                  3.00981900    3.61065300   -1.14063000
\end{verbatim}
T$_1$ geometry (\AA) of INVEST molecule (\textbf{1})
\begin{verbatim}
 C                  0.69384000    2.76547200   -0.63739800
 N                  1.64652800    1.86732500   -0.57404100
 C                  1.28893400    0.57653300   -0.39733000
 N                 -0.64747400    2.59096500   -0.56992400
 C                 -1.04203700    1.34631000   -0.39272000
 N                 -0.11678300    0.32112500   -0.28711500
 C                 -0.54385400   -1.01204100   -0.13540400
 C                 -2.71781600   -0.21988900   -0.12165300
 N                 -2.35111600    1.05690500   -0.31083300
 N                  2.13710600   -0.39860500   -0.32658000
 C                  1.69277400   -1.67676400   -0.14811500
 N                 -1.85315300   -1.26454200   -0.05394200
 N                  0.33218900   -1.98656300   -0.07155900
 C                  4.03009000   -2.69977600   -0.07364900
 C                  4.77020200   -3.86667400    0.04488700
 C                  4.20239400   -5.14449000    0.19707800
 C                  2.80182800   -5.23071600    0.22503900
 C                  2.02307400   -4.10104400    0.11123400
 C                  2.59156600   -2.79905200   -0.04021200
 H                  5.85292800   -3.78602800    0.01790800
 H                  2.32488700   -6.19911400    0.33546500
 H                  0.94638100   -4.18126400    0.13167400
 C                 -5.18778500    0.41228100    0.09603500
 C                 -4.14911600   -0.55340900    0.00358800
 C                 -4.49170500   -1.92187500    0.03003000
 C                 -5.80279900   -2.34414900    0.13165500
 C                 -6.84021900   -1.40919600    0.22069200
 C                 -6.50154300   -0.05610000    0.20263000
 H                 -3.69307800   -2.64724100   -0.03526700
 H                 -6.02584600   -3.40585700    0.14203900
 H                 -7.29758000    0.67787700    0.27632300
 C                  5.07120800   -6.36143200    0.35301500
 H                  5.96174100   -6.30184500   -0.27854100
 H                  4.52961900   -7.27485100    0.09797500
 H                  5.41876900   -6.46530800    1.38850400
 C                 -8.27422900   -1.85448900    0.34648800
 H                 -8.53267100   -2.57548000   -0.43445000
 H                 -8.44614400   -2.34738800    1.30910800
 H                 -8.96438200   -1.01232400    0.27384900
 C                 -4.98525500    1.90854000    0.10221700
 H                 -4.29423000    2.22362400    0.88598500
 H                 -4.54970100    2.26310200   -0.83300100
 H                 -5.94342100    2.40781100    0.25862900
 C                  4.77819600   -1.40018300   -0.22862800
 H                  5.85373200   -1.58866500   -0.20994700
 H                  4.52311800   -0.89400400   -1.16174400
 H                  4.52933200   -0.69216100    0.56452500
 O                  1.01432700    4.06159600   -0.81732200
 C                  2.77548000    5.06070800    0.49095800
 F                  4.08620000    5.38557200    0.46197200
 F                  2.08096100    6.18401300    0.74521000
 F                  2.57750900    4.22003500    1.52187300
 C                  2.38776400    4.43565900   -0.83866500
 H                  2.50123400    5.18951500   -1.61699800
 H                  3.03163000    3.57863500   -1.02762400 
\end{verbatim}
S$_0$ geometry (\AA) of non-INVEST molecule (\textbf{2})
\begin{verbatim}
 C                  0.00005100    2.87398300   -0.00014100
 N                  1.20635000    2.26172400    0.01699900
 C                  1.21620000    0.94204400    0.00771200
 N                 -1.20626400    2.26177300   -0.01728600
 C                 -1.21617500    0.94209100   -0.00789800
 N                  0.00000100    0.23954400   -0.00004400
 C                 -0.00002700   -1.16446700    0.00000600
 C                 -2.29198000   -1.08370300   -0.01004700
 N                 -2.35709700    0.25327400   -0.00808800
 N                  2.35709900    0.25317600    0.00792100
 C                  2.29192600   -1.08379300    0.01001800
 N                 -1.15825100   -1.80965300   -0.00762300
 N                  1.15816300   -1.80970200    0.00769000
 C                  4.85286400   -1.27439500    0.09112000
 C                  5.95453200   -2.13801200    0.07385100
 C                  5.84672100   -3.52485600   -0.02230000
 C                  4.56384400   -4.07605800   -0.09768600
 C                  3.45174800   -3.25496800   -0.08071500
 C                  3.55918700   -1.85349000    0.00863700
 H                  6.94513200   -1.69913200    0.14154300
 H                  4.43749500   -5.15154900   -0.16797600
 H                  2.45858700   -3.67849100   -0.13505400
 C                 -4.85293200   -1.27420700   -0.09097700
 C                 -3.55927400   -1.85334800   -0.00861200
 C                 -3.45188200   -3.25484000    0.08068900
 C                 -4.56400100   -4.07588900    0.09769100
 C                 -5.84686800   -3.52463600    0.02238500
 C                 -5.95463400   -2.13779000   -0.07367500
 H                 -2.45873300   -3.67839700    0.13497500
 H                 -4.43768800   -5.15138500    0.16796100
 H                 -6.94522300   -1.69886800   -0.14125100
 C                  7.07202100   -4.40260300   -0.05986300
 H                  7.96857400   -3.85297900    0.23298000
 H                  7.23777200   -4.79712600   -1.06821900
 H                  6.96440200   -5.26141400    0.60840100
 C                 -7.07217200   -4.40238500    0.05978600
 H                 -6.96540900   -5.25989700   -0.61029800
 H                 -7.23668600   -4.79889000    1.06755900
 H                 -7.96905300   -3.85216600   -0.23091600
 C                 -5.14303700    0.20392300   -0.20609800
 H                 -4.77169800    0.75685300    0.65760600
 H                 -4.65002500    0.64821200   -1.07178300
 H                 -6.22068700    0.36000500   -0.29258400
 C                  5.14299500    0.20371900    0.20639100
 H                  6.22063500    0.35976200    0.29306300
 H                  4.77181700    0.75671200   -0.65734500
 H                  4.64985100    0.64796900    1.07201900
 C                  1.05120500    6.32453200   -0.72000400
 C                  0.00018400    7.19705200    0.00023100
 C                 -1.05092500    6.32440400    0.72018700
 C                 -1.26241600    4.98203800    0.02042300
 H                  0.73658100    6.12903500   -1.75013300
 H                 -0.48690400    7.85249500   -0.72744700
 H                 -2.00996300    6.84470600    0.78332000
 H                 -1.61782200    5.13392000   -1.00683600
 N                  0.00009000    4.22090000   -0.00014900
 C                  1.26264900    4.98196000   -0.02062800
 H                  2.01172600    4.37822400   -0.52741400
 H                  1.61816000    5.13352800    1.00664000
 H                  2.01027400    6.84479000   -0.78303000
 H                  0.48733600    7.85221700    0.72811700
 H                 -0.73635100    6.12858700    1.75027100
 H                 -2.01156800    4.37821500    0.52698800
\end{verbatim}
S$_1$ geometry (\AA) of non-INVEST molecule (\textbf{2})
\begin{verbatim}
 C                 -0.02768000    2.88738900   -0.10346700
 N                  1.17724000    2.27641600   -0.20344600
 C                  1.20851100    0.95302700   -0.12433300
 N                 -1.21684600    2.25219800   -0.22383600
 C                 -1.22237500    0.92749600   -0.13740800
 N                  0.00081700    0.24303400    0.10439500
 C                  0.01654800   -1.14518300   -0.22994500
 C                 -2.29153000   -1.11986100   -0.19258900
 N                 -2.33560000    0.24067300   -0.26369800
 N                  2.34040000    0.29272600   -0.24074800
 C                  2.32470000   -1.06960900   -0.17316000
 N                 -1.12343100   -1.79396900   -0.36013500
 N                  1.17796900   -1.76746600   -0.35308700
 C                  4.83276300   -1.23765200    0.25395700
 C                  5.94153300   -2.08333200    0.34221200
 C                  5.87778200   -3.46418800    0.14859200
 C                  4.62685700   -4.02261900   -0.13891400
 C                  3.50632600   -3.22062100   -0.23381600
 C                  3.57120500   -1.82070800   -0.05187400
 H                  6.90261300   -1.63764000    0.58060400
 H                  4.53341000   -5.09323300   -0.29039800
 H                  2.54350700   -3.65818100   -0.45943200
 C                 -4.80038200   -1.34507700    0.19888200
 C                 -3.51870300   -1.89961700   -0.07955100
 C                 -3.41661200   -3.30065300   -0.23998000
 C                 -4.51772700   -4.12937400   -0.14882200
 C                 -5.78665800   -3.59926000    0.11218300
 C                 -5.88809100   -2.21745000    0.28357000
 H                 -2.44013200   -3.71705200   -0.44577100
 H                 -4.39482300   -5.19938500   -0.28280000
 H                 -6.86378600   -1.79300800    0.50060200
 C                  7.11365100   -4.32348400    0.22907700
 H                  7.93733400   -3.79620300    0.71462700
 H                  7.45266800   -4.61763100   -0.77064500
 H                  6.92230500   -5.24397600    0.78763200
 C                 -7.00165400   -4.48826300    0.18710700
 H                 -7.32426800   -4.79740900   -0.81359800
 H                 -6.79319200   -5.40003800    0.75376900
 H                 -7.84266500   -3.97817300    0.66129200
 C                 -5.06423000    0.12391200    0.43285300
 H                 -4.44383200    0.52530300    1.23757200
 H                 -4.83711000    0.72404500   -0.44931400
 H                 -6.11248200    0.27200700    0.70167700
 C                  5.05500000    0.23379800    0.51472500
 H                  6.09297400    0.40342100    0.80929000
 H                  4.83345300    0.84062400   -0.36431100
 H                  4.40745200    0.60860800    1.31070700
 C                  0.95982500    6.35532800   -0.70056200
 C                 -0.13764200    7.20302800   -0.01694600
 C                 -1.13039400    6.31459700    0.76092300
 C                 -1.33082700    4.95302900    0.09692600
 H                  0.67709200    6.13622300   -1.73492100
 H                 -0.66965000    7.78639600   -0.77393600
 H                 -2.10085300    6.80868300    0.85303200
 H                 -1.75121800    5.07030400   -0.91071600
 N                 -0.04772600    4.23468700    0.02486500
 C                  1.19274000    5.02608300    0.01957200
 H                  1.96439300    4.43861500   -0.47248500
 H                  1.52880000    5.20123200    1.05020100
 H                  1.90357400    6.90485000   -0.74605500
 H                  0.31257200    7.92603500    0.66918300
 H                 -0.76814200    6.15123500    1.78105300
 H                 -2.03736600    4.34186000    0.65770400
\end{verbatim}
T$_1$ geometry (\AA) of non-INVEST molecule (\textbf{2})
\begin{verbatim}
 C                 -0.01566300    2.88870900   -0.09445000
 N                 -1.21452400    2.26365100   -0.17434700
 C                 -1.22228200    0.93774500   -0.10694500
 N                  1.19380900    2.27584700   -0.15021200
 C                  1.21212800    0.95207500   -0.09483500
 N                  0.00116200    0.25697100    0.06001300
 C                  0.00951100   -1.15463600   -0.15138200
 C                  2.32067200   -1.08339200   -0.10591400
 N                  2.35065300    0.28635000   -0.17419200
 N                 -2.34641000    0.25685400   -0.19226300
 C                 -2.30154800   -1.11278400   -0.12258900
 N                  1.17171800   -1.78655800   -0.22104100
 N                 -1.13881400   -1.80158000   -0.22861200
 C                 -4.81934700   -1.33139400    0.23021700
 C                 -5.91777700   -2.19456300    0.26581600
 C                 -5.82910800   -3.56972400    0.04316800
 C                 -4.56147100   -4.10275300   -0.21717600
 C                 -3.45049400   -3.28289100   -0.25999200
 C                 -3.53971000   -1.88750500   -0.05077700
 H                 -6.89172800   -1.76767300    0.48629600
 H                 -4.44667000   -5.16835400   -0.38876600
 H                 -2.47534000   -3.70370800   -0.46276900
 C                  4.83546600   -1.26817000    0.28662400
 C                  3.57132400   -1.84047600   -0.02536600
 C                  3.50722800   -3.23251900   -0.25916400
 C                  4.63064300   -4.03521300   -0.21188300
 C                  5.88469500   -3.48607700    0.07881500
 C                  5.94752600   -2.11407000    0.32648800
 H                  2.54221100   -3.66502800   -0.48481700
 H                  4.53629500   -5.09935200   -0.40354700
 H                  6.91040700   -1.67559200    0.57125900
 C                 -7.05460500   -4.44740000    0.06535400
 H                 -7.89668200   -3.94718600    0.54824600
 H                 -6.86425400   -5.38345000    0.59781900
 H                 -7.36677900   -4.71251900   -0.95120200
 C                  7.12344700   -4.34479600    0.10788500
 H                  6.94303300   -5.28559100    0.63546900
 H                  7.44742900   -4.60138200   -0.90705200
 H                  7.95367700   -3.83354800    0.59949400
 C                  5.05878500    0.19117400    0.61011800
 H                  4.85715900    0.83398500   -0.24778300
 H                  4.39719000    0.53749600    1.40735500
 H                  6.09170700    0.34326500    0.93114000
 C                 -5.07223100    0.12944700    0.52359000
 H                 -6.11545800    0.27122700    0.81480900
 H                 -4.43667900    0.49778900    1.33214900
 H                 -4.85665200    0.76183900   -0.33862800
 C                 -1.09143600    6.33572900    0.69740700
 C                 -0.08791600    7.20703100   -0.08613500
 C                  0.99510300    6.34000400   -0.76745400
 C                  1.22145100    5.01809200   -0.03266400
 H                 -0.73426600    6.17849900    1.72025800
 H                  0.37453300    7.92619300    0.59580000
 H                  1.94414600    6.87873000   -0.82693800
 H                  1.56032600    5.20118900    0.99518500
 N                 -0.02672700    4.23545700   -0.01303100
 C                 -1.30377400    4.96992600    0.04607400
 H                 -2.01850100    4.37030000    0.60735000
 H                 -1.71427900    5.08115200   -0.96580000
 H                 -2.05767700    6.83921900    0.78190500
 H                 -0.61254100    7.79474700   -0.84481900
 H                  0.70263900    6.11133300   -1.79708100
 H                  1.98697000    4.41806900   -0.51850200   
\end{verbatim}
\bibliography{manuscript}% Produces the bibliography via BibTeX.